%% file: main.tex
\documentclass[reprint, superscriptaddress, amsmath, amssymb, aps, prb]{revtex4-2}

\usepackage{graphicx}
\usepackage{dcolumn}
\usepackage{bm}
\usepackage{hyperref}

\usepackage[dvipsnames]{xcolor}

\newcommand{\veck}{\textbf{k}}
\newcommand{\veckP}{\textbf{k}^\prime}

\def\={&=&}

\begin{document}

\preprint{APS/123-QED}

\title{Thermodynamics of $T_{\rm c}$ suppression in far-overdoped Tl$_2$Ba$_2$CuO$_6$}
\author{Ayanesh Maiti}
\affiliation{Max Planck Institute for Chemical Physics of Solids, Dresden, Germany.}
\affiliation{Max Planck Institute for Structure and Dynamics of Matter, Hamburg, Germany.}
\affiliation{SUPA School of Physics and Astronomy, University of St Andrews, North Haugh, St Andrews, KY16 9SS, United Kingdom.}

\author{David M. Broun}
\affiliation{Department of Physics, Simon Fraser University, Burnaby, British Columbia V5A 1S6, Canada}

\author{Seunghyun Khim}
\affiliation{Max Planck Institute for Chemical Physics of Solids, Dresden, Germany.}

\author{Michal Moravec}
\affiliation{Max Planck Institute for Chemical Physics of Solids, Dresden, Germany.}

\author{Antony Carrington}
\affiliation{H.H. Wills Physics Laboratory, University of Bristol, Tyndall Avenue, BS8 1TL, United Kingdom}

\author{Carsten Putzke}
\affiliation{Max Planck Institute for Structure and Dynamics of Matter, Hamburg, Germany.}

\author{Vivek Mishra}
\affiliation{Department of Physics, University of Florida, Gainesville, Florida 32611, USA}

\author{Peter J. Hirschfeld}
\affiliation{Department of Physics, University of Florida, Gainesville, Florida 32611, USA}

\author{Andrew P. Mackenzie}
\affiliation{Max Planck Institute for Chemical Physics of Solids, Dresden, Germany.}
\affiliation{SUPA School of Physics and Astronomy, University of St Andrews, North Haugh, St Andrews, KY16 9SS, United Kingdom.}
\email{apm9@st-andrews.ac.uk}

\author{Andreas W. Rost}
\affiliation{Max Planck Institute for Chemical Physics of Solids, Dresden, Germany.}
\affiliation{SUPA School of Physics and Astronomy, University of St Andrews, North Haugh, St Andrews, KY16 9SS, United Kingdom.}
\email{a.rost@st-andrews.ac.uk}

\date{\today}

\begin{abstract}
The physical origin of the suppression of superconductivity with hole doping in overdoped cuprates remains unclear. We measure the electronic specific heat of \textmu{}g-scale Tl$_2$Ba$_2$CuO$_6$ crystals and find sharp superconducting anomalies persisting far into the overdoped regime. A weak-coupling BCS-like framework incorporating the known Fermi surface and cation disorder quantitatively reproduces the observed anomalies for $T_{\rm c}=14$--25\,K and their weak doping dependence. The results show $T_{\rm c}(p)$ to be driven predominantly by a smoothly decreasing pairing strength.
\end{abstract}

\maketitle

\begin{figure*}
    \centering
    \includegraphics[scale=0.84]{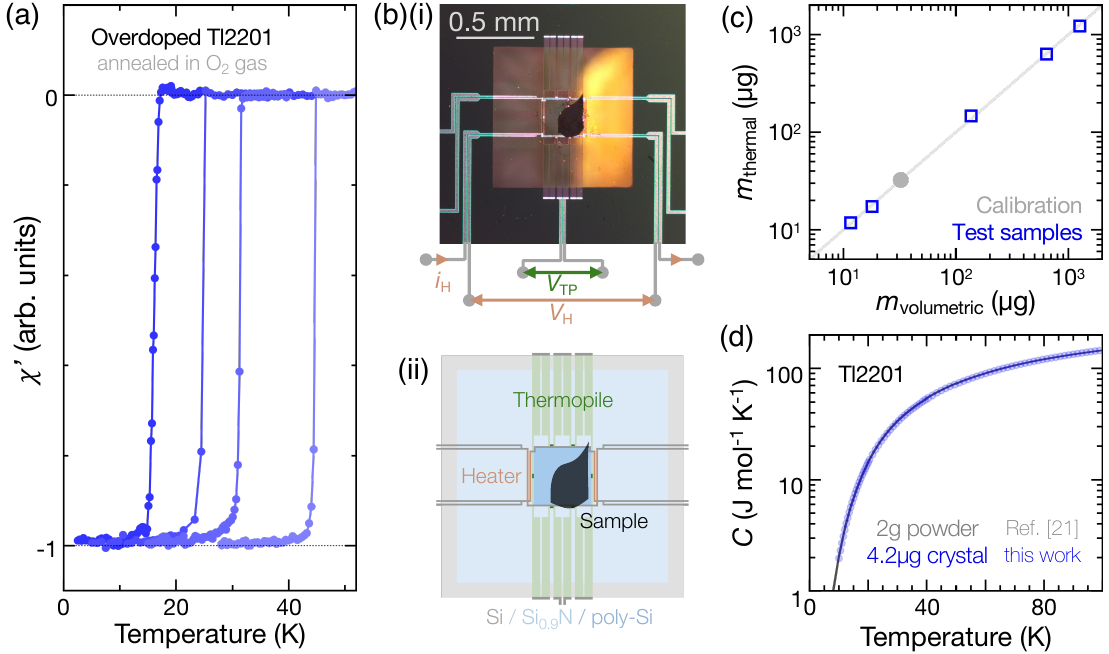}
    \caption{{\bf Specific heat measurements on microgram-scale single crystals of Tl$_2$Ba$_2$CuO$_6$.}
    (a) AC magnetic susceptibility of overdoped Tl$_2$Ba$_2$CuO$_6$ single crystals with $T_c=14$--45\,K, used for the specific heat measurements. All samples exhibit a single, sharp superconducting transition with $\Delta T_c\sim1$\,K, indicative of homogeneous doping.
    (b) (i) Optical image of a 2~\textmu{}g Tl$_2$Ba$_2$CuO$_6$ single crystal mounted on a XEN-39398 nanocalorimeter chip. (ii) Schematic of the calorimeter, consisting of a Si$_{0.9}$N membrane incorporating a resistive heater and a thermopile. The central region is coated with a poly-Si thermalization layer, on which the sample is mounted using Apiezon N grease.
    (c) Validation of the AC nanocalorimetry technique over the temperature range 2--100\,K using the reference samples described in the main text. The sample mass inferred from the measured heat capacity and known specific heat is compared with the mass estimated from the sample dimensions and density. The two estimates agree to within 5\% over the mass range 2--2000\,\textmu{}g.
    (d) Temperature dependence of the specific heat of Tl$_2$Ba$_2$CuO$_6$ measured on a 4.2\,\textmu{}g single crystal, showing less than 2\% deviation from literature data on 2\,g polycrystalline samples.}
    \label{fig1}
\end{figure*}

\begin{figure*}
    \centering
    \includegraphics[scale=0.84]{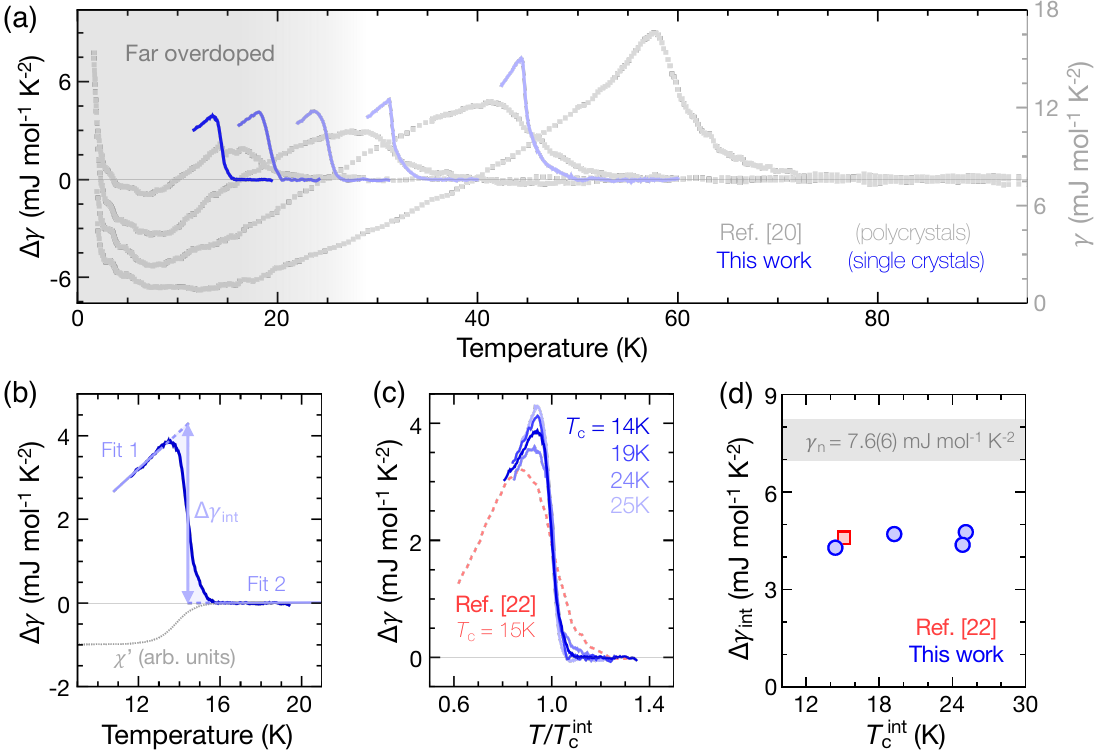}
    \caption{{\bf Robust superconducting transition in the electronic specific heat of far-overdoped Tl$_2$Ba$_2$CuO$_6$.}
    (a) Electronic specific heat of Tl$_2$Ba$_2$CuO$_6$ single crystals measured in this work (blue shades) in the vicinity of the superconducting transition, compared with literature data on polycrystalline samples (gray). In the far-overdoped regime ($T_{\rm c}<30$\,K), the single crystals exhibit sharp, well-resolved superconducting anomalies, whereas the polycrystal data show substantially broader and smaller anomalies.
    (b) Electronic specific heat anomaly of the most overdoped sample. The broadened transition is consistent with the transition width measured by AC magnetic susceptibility (gray). The intrinsic $T_{\rm c}^{\rm int}$ and $\Delta \gamma_{\rm int}$ are estimated by linear extrapolation of the superconducting and normal-state heat capacities to an intermediate temperature chosen to conserve entropy, as illustrated.
    (c) Electronic specific heat anomalies of the far-overdoped samples plotted as a function of $T/T_{\rm c}^{\rm int}$. The normalized anomalies are similar across the doping range studied, with only modest differences in transition broadening. Literature data on Tl$_2$Ba$_2$CuO$_6$ single crystals show a similar anomaly despite a broader transition.
    (d) Intrinsic $\Delta \gamma_{\rm int}$, extracted from the far-overdoped Tl$_2$Ba$_2$CuO$_6$ single crystals. The best available estimate of $\gamma_{\rm n}$, discussed in the text, is marked by the gray band. $\Delta \gamma_{\rm int}$ remains nearly independent of doping over the range $T_{\rm c}=14$--25\,K, with $\Delta \gamma_{\rm int}/\gamma_{\rm n}\approx0.59\pm0.05$.}
    \label{fig2}
\end{figure*}

\begin{figure*}
    \centering
    \includegraphics[scale=0.84]{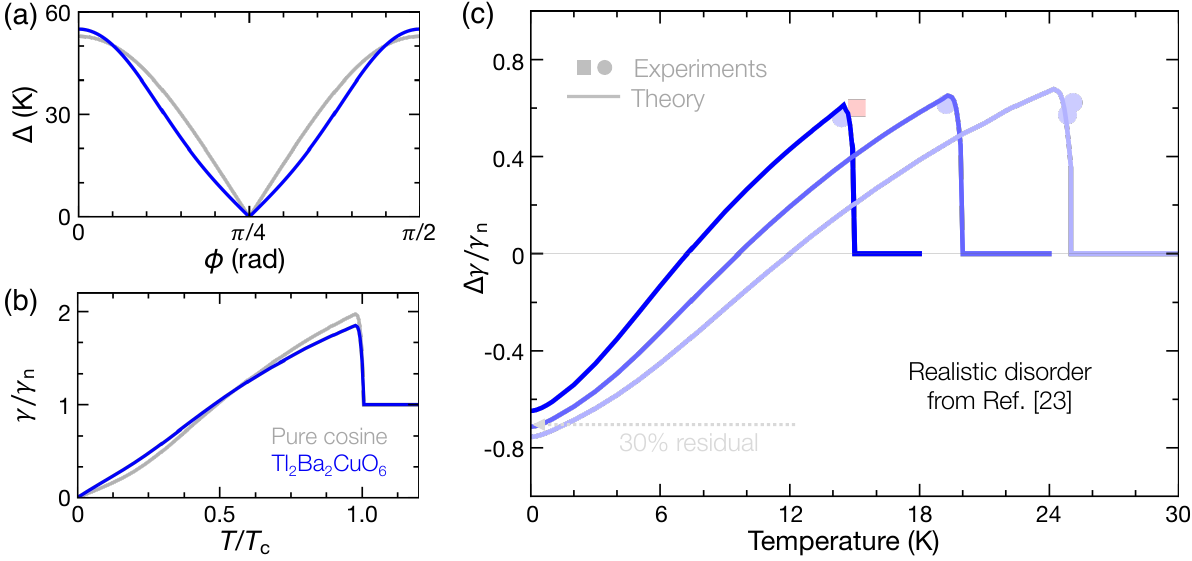}
    \caption{{\bf Realistic model of the electronic specific heat anomaly of far-overdoped Tl$_2$Ba$_2$CuO$_6$.} (a) $d$-wave superconducting gap on the Fermi surface of Tl$_2$Ba$_2$CuO$_6$, assuming a pure cosine form (gray) and a modulated form containing 12\% of the second harmonic (blue). (b) Calculated specific heat anomalies in the clean limit corresponding to the gap functions shown in panel (a), assuming a mean-field BCS-like description of the superconductivity in far-overdoped Tl$_2$Ba$_2$CuO$_6$. The jump at $T_{\rm c}$ is slightly suppressed by the higher harmonic content. (c) Calculated electronic specific heat anomalies at $T_{\rm c}=15$, 20, 25\,K in the presence of realistic levels of disorder as described in the text. The calculated curves exhibit a residual $\gamma_0/\gamma_{\rm n}\approx30\%$, and their specific heat jumps agree quantitatively with the experimentally derived $\Delta \gamma_{\rm int}$ values.}
    \label{fig3}
\end{figure*}

Although much has been learned in the forty years since the discovery of high-temperature superconductivity in the cuprates~\cite{Bednorz1986}, there is still no consensus on its microscopic mechanism~\cite{Keimer2015}.  The phase diagram is complex, with competing and intertwined orders making it difficult to identify a well-defined set of experimental facts against which microscopic theories can be tested. The strongly overdoped regime offers a particularly promising setting: the normal state shows many characteristics of a conventional Fermi liquid, while the superconducting state retains a number of standard features. However, even here there is controversy.  For example, the doping dependence of the superfluid density in La$_{2-x}$Sr$_x$CuO$_4$ has been argued to be both unconventional~\cite{Bozovic2018,Juskus2024} and conventional~\cite{Tallon2026a, Tallon2026}, and the potential role of disorder in cloaking the intrinsic behaviour has been stressed by a number of authors~\cite{Lee-Hone2017,Lee-Hone2020,Ozdemir2022,Ramshaw2026}.

One of the most attractive materials to study the strongly overdoped limit is Tl$_2$Ba$_2$CuO$_6$, which has one of the largest interplane spacings ($\sim11.7$\,\AA), placing the cation disorder well away from the CuO$_2$ planes. High-quality single crystals with superconducting transition temperature $T_{\rm c}<30$\,K exhibit some of the longest normal-state resistive mean free paths ($\sim500$\,\AA) observed in any cuprate~\cite{Shimakawa1990,Shimakawa1989,Kubo1991,Mackenzie1993,Mackenzie1996}. Quantum oscillations have been reported in such crystals~\cite{Vignolle2008, Rourke2010,Bangura2010}, revealing a large, quasi-two-dimensional Fermi surface containing $1+p$ holes, where $p$ is the effective hole doping of the CuO$_2$ layers. These properties make Tl$_2$Ba$_2$CuO$_6$ an unusually clean and well-characterized setting in which to investigate the superconducting state deep in the overdoped regime.

However, the existing heat capacity data tell a very different story. The only systematic calorimetry study reported to date~\cite{Wade1994,Loram1994,Wade1995} found broad superconducting transitions, and the residual electronic specific heat estimated in the most overdoped dataset reached $70\pm10\%$ of the normal-state value~\cite{Lee-Hone2020}, suggesting that superconductivity in strongly overdoped Tl$_2$Ba$_2$CuO$_6$ is dominated by disorder and/or doping inhomogeneity over a range of length scales. These results were obtained on polycrystalline samples, motivating a systematic study of carefully characterized single crystals to distinguish the effects of doping inhomogeneity and disorder on the superconducting thermodynamics. Subsequently, an AC heat capacity measurement on a single crystal with $T_{\rm c}\sim15$\,K showed a sharper transition and larger specific heat anomaly than the polycrystalline data~\cite{Carrington1996}, further emphasizing the need for a systematic study of small, highly homogeneous single crystals. Here we report such a study and show that the thermodynamics of far-overdoped single-crystal Tl$_2$Ba$_2$CuO$_6$ differs markedly from that inferred from earlier polycrystalline measurements~\cite{Wade1994,Wade1995,Loram1994}. In particular, we observe sharp, reproducible superconducting anomalies deep in the overdoped regime and use their magnitude to constrain the role of disorder and the underlying pairing strength.

The single crystals used in this study were grown following the self-flux method described in Refs.~\cite{Liu1992,Mackenzie1993,Tyler1997, Putzke2021}.  The samples were annealed at 450$^\circ$C for 14--18\,hrs in selected static O$_2$ pressures between 0.001 and 1\,bar to vary $T_{\rm c}$.  As illustrated in Fig.~\ref{fig1}a, all samples exhibit sharp superconducting transitions in magnetic susceptibility, with $\Delta T_{\rm c}\sim1$\,K over the range $T_{\rm c}=14$--45\,K.

Single crystals of Tl$_2$Ba$_2$CuO$_6$ with sharp susceptibility transitions typically have a mass in the 1--10\,\textmu{}g range.  To measure the heat capacity of such small samples, we built on established membrane nanocalorimetry techniques~\cite{Tagliati2011thesis,Campanini2019,Tagliati2011b,Tagliati2012,Walmsley2014,Walmsley2013,Girod2020,Cole2025}. Specifically, we used XENSOR XEN-39398 nanocalorimeter chips, which incorporate a poly-Si heater and thermopile (six thermocouples in series) on a Si$_{0.9}$N membrane, with a central sample region coated with a poly-Si thermalization layer.  A photograph and schematic of the chip with a 2\,\textmu{}g single crystal of Tl$_2$Ba$_2$CuO$_6$, mounted with approximately 2--4\,ng of Apiezon N grease, are shown in Fig.~\ref{fig1}b.  AC measurements were conducted in the low frequency limit following the procedures of Refs.~\cite{Tagliati2011thesis,Campanini2019}, and using the quasi-adiabatic, quasi-static technique of Refs.~\cite{Walmsley2014,Walmsley2013}. The measured data were corrected for the addenda contribution from the calorimeter chip (see Supplementary Fig.~\ref{figS1}c).

The above-described setup provides sufficient sensitivity to measure the heat capacity of \textmu{}g-scale crystals, but considerable care is required to obtain heat capacity in absolute units because the thermopiles are not calibrated by the manufacturer.  We determined the effective thermopower by performing nanocalorimetry measurements on a known mass of high purity Ag, using tabulated specific heat data from NIST~\cite{Smith1995}. The calibration was then validated by measuring samples of Sr$_2$RuO$_4$, Au, Pb, NbSe$_2$, and YBa$_2$Cu$_3$O$_{6.67}$.  As shown in Fig.~\ref{fig1}c, the ‘thermal’ mass $m_{\rm thermal}$ (deduced from the measured heat capacity and known specific heat~\cite{Wade1995,Tagliati2012Pb,Pelly2026,Geballe1952,Martin1987,Loram1994ybco,Mackenzie1998b}) agrees quantitatively with the ‘volumetric’ mass $m_{\rm volumetric}$ (deduced from the measured volume of each sample and its known density).  The two values agree within 5\% for all samples, with the uncertainty dominated by the determination of sample volume.

Finally, we show in Fig.~\ref{fig1}d a comparison of the total specific heat of a 4.2\,\textmu{}g single crystal of Tl$_2$Ba$_2$CuO$_6$ with that reported for a 2\,g polycrystalline sample~\cite{Wade1995}. To account for uncertainty in the sample mass, the specific heat data was scaled by a factor of 1.05 to agree with the literature dataset at the highest temperature, following a procedure similar to that used in Ref.~\cite{Carrington1996}. The resulting curves agree within 1--2\% over the entire temperature range (10--100\,K) relevant to this study, providing a stringent check of the measured temperature dependence. We obtain comparable agreement for sample masses in the range 1–100\,\textmu{}g, with scaling factors in the range 0.95--1.05.

In common with other cuprates, the heat capacity near $T_{\rm c}$ is dominated by the phonon contribution, with the superconducting anomaly changing the total specific heat by less than 1\%. To isolate the electronic contribution, we fit the normal-state specific heat above $T_{\rm c}$ with a fourth-order polynomial, extrapolate the fit to 5--10 K below $T_{\rm c}$, and subtract it from the total heat capacity, following procedures used in Refs.~\cite{Walmsley2013,Walmsley2014} (see End Matter, Supplementary Table~\ref{table1}). This procedure was verified against measurements in high magnetic fields where the effects of superconductivity are suppressed. However, this procedure can still introduce systematic errors that increase with the extrapolation range. We therefore restrict our analysis to the superconducting anomaly and a limited temperature range below $T_{\rm c}$, where we could ensure that the physical conclusions drawn in this work are independent of the background uncertainty.

In Fig.~\ref{fig2}a, we show the specific heat transitions for five single crystal measurements from this work (blue), alongside those from Ref.~\cite{Loram1994}. In all cases, we plot the change in the electronic specific heat coefficient $\Delta\gamma=\Delta C_{\rm el}/T$, with a moving average filter applied to reduce noise from the datasets. Unfiltered data are shown in Supplementary Materials Fig.~\ref{figS2}.

The transitions observed in all of the single crystals are substantially narrower than those in the polycrystals for the range of transition temperatures studied by us, while the jump magnitudes at comparable $T_{\rm c}$ are markedly larger.
For $T_{\rm c}>30$\,K, significant fluctuation contributions appear above $T_{\rm c}$  and increase towards optimal doping. The transition in the polycrystalline sample with $T_{\rm c}=57$\,K is dominated by fluctuations rather than disorder broadening. Given the $T_{\rm c}$ dependence of the fluctuation contributions, we assume them to be negligible for samples with $T_{\rm c}<25$\,K. Our background subtraction could, however, include a very broad fluctuation contribution indistinguishable from the phonon background; the extracted jump would then be a lower bound on the intrinsic value though we expect any such effect to be small.

As illustrated by the data for $T_{\rm c} = 14$\,K (Fig.~\ref{fig2}b), there is excellent agreement between the susceptibility and heat capacity transition widths for the same crystals \footnote{No susceptibility data is available for the crystal that was annealed to $T_{\rm c}=19$\,K.}. Strikingly, the shape and magnitude of the transitions are qualitatively similar across all single-crystal datasets with $T_{\rm c} \leq 25$\,K. To investigate these transitions quantitatively, we attribute the finite transition widths to residual doping inhomogeneity and impose entropy conservation across the transition to determine the intrinsic transition temperature $T_{\rm c}^{\rm int}$ and jump height $\Delta\gamma_{\rm int}$, as illustrated in Fig.~\ref{fig2}b.
  
In Fig.~\ref{fig2}c we show the transitions of four samples from the present study and one from a previous study~\cite{Carrington1996}, plotted as a function of reduced temperature $T/T_{\rm c}^{\rm int}$.  The transition widths from all four samples studied here are very similar as a function of reduced temperature, while the dataset from Ref.~\cite{Carrington1996} is broader, possibly due to less homogeneous doping. Notably, all of the plotted datasets look plausibly like disorder-broadened mean-field anomalies.  The jump height $\Delta\gamma_{\rm int}$ obtained using the procedure shown in Fig.~\ref{fig2}b is shown for all five samples in Fig.~\ref{fig2}d. Consistent with the similarity in the anomaly magnitudes observed above, we obtain a roughly constant $\Delta\gamma_{\rm int}=4.5\pm0.2$\,mJ\,mol$^{-1}$\,K$^{-2}$.

Our measurements do not provide a value for the normal state $\gamma_{\rm n}$, but independent estimates are available from two sources.  In the original work on polycrystals~\cite{Wade1994} an estimate of $\gamma_{\rm n}=7\pm1$\,mJ\,mol$^{-1}$\,K$^{-2}$ was given.  Later, a cyclotron mass of $m^* = (5.2\pm0.3)m_{e}$ was obtained from quantum oscillation measurements for $T_{\rm c}=10$--25\,K~\cite{Bangura2010,Rourke2010}. This effective mass corresponds to $\gamma_{\rm n}=7.6\pm0.6$\,mJ\,mol$^{-1}$\,K$^{-2}$.  In both cases, no $T_{\rm c}$ dependence of $\gamma_{\rm n}$ was resolved over the relevant range, 14\,$\textrm{K}<T_{\rm c} < 25$\,K.  Taking the more precise estimate from the quantum oscillation studies, we obtain $\Delta\gamma_{\rm int}/\gamma_{\rm n} = 0.59 \pm 0.05$.

Within weak-coupling BCS theory on a circular Fermi surface, a $d$-wave gap of the form $\Delta(\phi)=\Delta_0\cos2\phi$  gives $\Delta\gamma(T_{\rm c})/\gamma_{\rm n}=0.95$ in the clean limit~\cite{Won1994}. The experimentally observed specific heat jump is substantially smaller than this ideal value.  Conservation of the change in entropy between $T = 0$ and $T_\mathrm{c}$ means that a reduced $\Delta \gamma(T_{\rm c})$ implies the existence of low energy states in excess of those found in the ideal $d$-wave reference state.  As outlined in End Matter, we allow for the possibility that the $d$-wave gap has higher harmonic content of the form $\Delta(\phi)\approx\Delta_0(\cos2\phi+\alpha\cos6\phi)$, reducing the near-nodal gap slope, with a corresponding increase in the density of states at low energies. Importantly, thermal conductivity measurements on $T_{\rm c} = 15$~K Tl$_2$Ba$_2$CuO$_6$ \cite{Proust2002} imply such a higher harmonic term exists, but constrain its magnitude to $\alpha = 12$\%.  As seen in Fig.~\ref{fig3}a--b, the corresponding heat capacity jump is indeed reduced, to $\Delta\gamma(T_{\rm c})/\gamma_{\rm n}=0.84$.  However, this is insufficient to account for the experimentally observed jump, implying an additional source of low energy states.

Elastic-scattering disorder provides a natural mechanism for forming the missing low energy states, as it breaks pairs and introduces zero-energy quasiparticles. To test this idea, we use the dirty $d$-wave framework of Refs.~\cite{Lee-Hone2017,Lee-Hone2020,Ozdemir2022}, as described in End Matter. The model is built on a semi-realistic parameterization of the  Tl$_2$Ba$_2$CuO$_6$ ARPES dispersion \cite{Plate2005,Peets2007b}, renormalized to agree with the observed cyclotron mass \cite{Bangura2010,Rourke2010}. This is coupled with an \textit{ab-initio} calculation of the impurity potential of the known out-of-plane cation disorder, an approximately 15\% Cu excess that substitutes onto the Tl sites in tetragonal Tl$_2$Ba$_2$CuO$_6$ crystals~\cite{Liu1992,Tyler1997,Maiti2026}, along with a small amount of strong scattering disorder. As shown in Fig.~\ref{fig3}c, the calculations reproduce both the absolute magnitude of the measured heat-capacity jumps and their weak dependence on $T_{\rm c}$ over the range 14--25\,K.  In this way, the measured thermodynamic anomalies can be quantitatively described by a weak-coupling \mbox{$d$-wave} framework incorporating the known level of cation disorder, without requiring a strongly doping-dependent increase in elastic scattering. Indeed, if the suppression of $T_{\rm c}$ were instead dominated by a hypothetical disorder level that increased with doping, the calculated heat-capacity jump would acquire a strong $T_{\rm c}$ dependence, contrary to experiment, as shown in End Matter.

While it is not our intention to perform a detailed fit of the heat-capacity anomalies, the calculations provide a useful benchmark for the role of disorder in these Tl$_2$Ba$_2$CuO$_6$ crystals. As indicated in Fig.~\ref{fig3}c, the expected residual electronic specific heat based on the model is approximately 30\% of the normal-state value, in marked contrast to the 70\% residual term expected from the earlier polycrystalline measurements~\cite{Wade1994,Wade1995,Loram1994,Lee-Hone2020}, but comparable in magnitude to the residual electronic contribution inferred from thermal conductivity of single crystals~\cite{Proust2002}. Thus, disorder has a significant, but not dominant, effect on the superconducting state.

The disorder model also provides a route to estimating the underlying dimensionless pairing strength, $N(0)V_p$, as described in End Matter.  This yields a smoothly decreasing pairing strength across the strongly overdoped regime. Our results therefore support a picture in which the low-$T_{\rm c}$ superconductivity of Tl$_2$Ba$_2$CuO$_6$ is broadly consistent with a weak-coupling BCS-like framework~\cite{Lee-Hone2017,Lee-Hone2020,Ozdemir2022}: the intrinsic superconducting scale is suppressed by a smoothly decreasing pairing strength, while fixed chemical disorder reduces the thermodynamic response, which varies only weakly with doping. The inferred pairing strength remains substantial at the overdoped edge of the observed superconducting dome in Tl$_2$Ba$_2$CuO$_6$, suggesting that rather than terminating abruptly at $p\approx30\%$, the intrinsic superconducting phase should have a persistent tail in the absence of disorder, consistent with previous proposals~\cite{Maier2020,Ramshaw2026}. The resulting doping dependence of the pairing strength provides a new benchmark for microscopic theories of cuprate superconductivity.

\vspace{12pt}
\textit{Acknowledgements --} We thank B.H.\,Goodge, S.A.\,Kivelson, and B.J.\,Ramshaw for helpful discussions. A.M.\,acknowledges funding from the Max Planck Society via the MPGC-QM and IMPRS-CPQM programmes. A.W.R.\,acknowledges support from the Engineering and Physical Sciences Research Council (grants EP/P024564/1 and EP/V049410/1). P.J.H.\,was supported by NSF-DMR-2231821. D.M.B.\,acknowledges funding from the Natural Sciences and Engineering Research Council of Canada. Research in Dresden benefits from the environment provided by the DFG Cluster of Excellence ctd.qmat (EXC2147, Project ID 390858490).

\vspace{6pt}
\textit{Data availability --} The data supporting the findings of this study will be made available in a public repository upon publication.

\bibliographystyle{apsrev4-2}
\bibliography{ref}

\input{endmatter}

\input{supplement}

\end{document}

%% file: endmatter.tex
\onecolumngrid
\section*{End Matter}
\twocolumngrid
\setcounter{figure}{0}
\renewcommand{\thefigure}{A\arabic{figure}}

\begin{figure}[]
    \centering
    \includegraphics[scale=0.78]{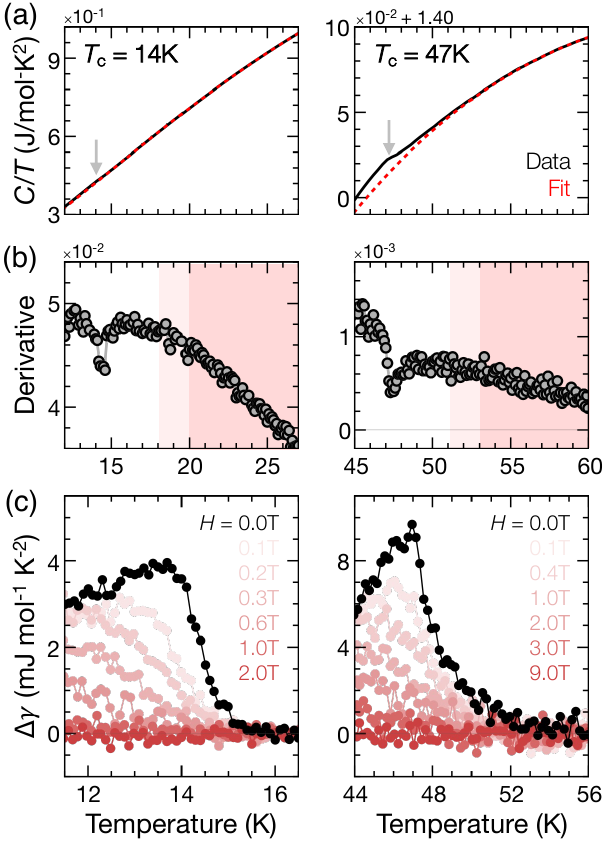}
    \caption{{\bf Polynomial background fits.}
    (a) Total specific heat data for samples with $T_{\rm c}=14$ and $47$\,K. Grey arrows indicate the values of $T_{\rm c}$ determined from AC magnetic susceptibility measurements. The data above $T_{\rm c}$ are fitted with fourth-order polynomials, which are extrapolated to estimate the normal-state background around $T_{\rm c}$.
    (b) Smoothed derivatives of the data from panel a. The red shaded regions mark the temperature ranges above $T_{\rm c}$ chosen for the background polynomial fits, where the derivative is approximately linear and featureless. The sensitivity of the background-subtracted datasets to the choice of lower cutoff temperature within the lighter red shaded region is shown in Fig.~\ref{figA2}.
    (c) Measurements on the same samples with a magnetic field applied along the $c$ axis. The polynomial backgrounds determined from the zero-field data are subtracted from all datasets. At the highest measured fields, the residual is featureless over the temperature ranges shown, providing an independent check of the chosen polynomial backgrounds.}
    \label{figA1}
\end{figure}

\begin{figure}[]
    \centering
    \includegraphics[scale=0.78]{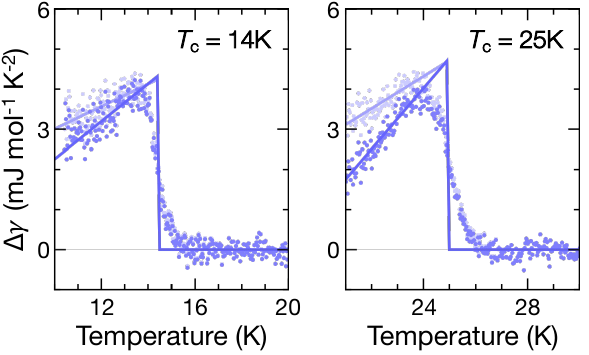}
    \caption{{\bf Sensitivity to background subtraction.}
    Two estimates of the electronic specific heat for the $T_{\rm c}=14$ and 25\,K datasets, obtained by varying the range of data and the order of the polynomial used to estimate the normal-state background. The uncertainty in the absolute electronic specific heat increases with the range of extrapolation below $T_{\rm c}$. In contrast, the intrinsic transition temperature $T_{\rm c}^{\rm int}$ and jump $\Delta\gamma_{\rm int}$ obtained from the entropy-balance construction vary by only 1--2\%.
    }
    \label{figA2}
\end{figure}

\begin{figure}[]
    \centering
    \includegraphics[scale=0.78]{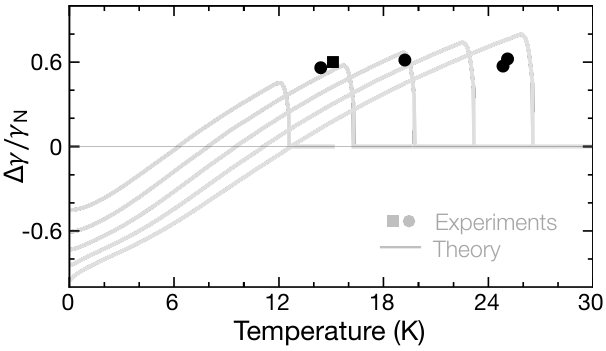}
    \caption{{\bf Calculations with increasing disorder.}
    Hypothetical electronic specific heat anomalies for $T_{\rm c}=15$--25\,K, calculated assuming a fixed pairing strength corresponding to clean-limit $T_\mathrm{c} = 30$~K and suppressing $T_{\rm c}$ solely by increasing the disorder. In this case, $\Delta\gamma(T_{\rm c})/\gamma_{\rm n}$ is predicted to decrease by approximately $40\%$ over the range of $T_{\rm c}$ studied experimentally. This is inconsistent with the approximately constant anomaly observed experimentally, for which the absolute uncertainty is $\sim10\%$. Thus, the observed doping dependence of $T_{\rm c}$ cannot be explained predominantly by an increase in disorder with doping.}
    \label{figA3}
\end{figure}

\begin{figure}[]
    \centering
    \includegraphics[scale=0.78]{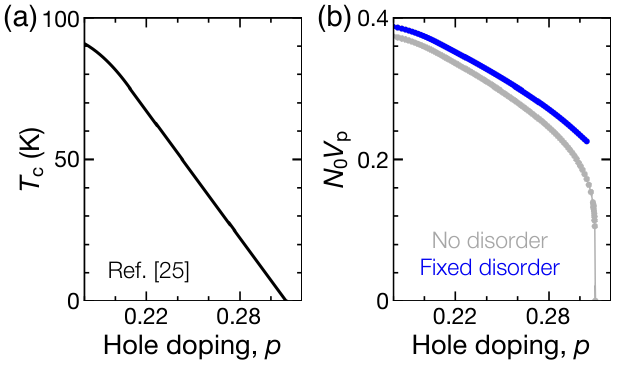}
    \caption{{\bf Inferred pairing strength.}
    (a) Approximate doping dependence of $T_{\rm c}$ in overdoped Tl$_2$Ba$_2$CuO$_6$, as determined by previous studies~\cite{Rourke2010,Bangura2010,Putzke2021}.
    (b) Pairing strength $N(0)V_p$ inferred using the realistic disorder model discussed in the main text. The pairing strength decreases smoothly with hole doping, accounting for the observed suppression of $T_{\rm c}$ without requiring an increase in disorder. By contrast, interpreting the observed $T_{\rm c}(p)$ directly as the clean-limit transition temperature (grey) would require a strongly nonlinear evolution of the pairing strength, with infinite slope as $T_{\rm c}\rightarrow0$.}
    \label{figA4}
\end{figure}

\subsection{Background subtraction}

The datasets shown in the main text are obtained by subtracting fourth-order polynomial backgrounds fitted over a temperature range above $T_{\rm c}$ where the derivative is approximately linear and featureless, as illustrated in Fig.~\ref{figA1}a–b. The resulting backgrounds are also consistent with high-field measurements, as illustrated in Fig.~\ref{figA1}c. Notably, as illustrated in Fig.~\ref{figA2}, uncertainties associated with this background subtraction procedure have little influence on the extracted values of $\Delta\gamma_{\rm int}$ and $T_{\rm c}^{\rm int}$, which are the quantities discussed in the main text.

\subsection{Dirty $d$-wave theory with realistic disorder}

To capture the effect of the dominant source of disorder in Tl$_2$Ba$_2$CuO$_6$ --- an approximately 15\% excess of Cu atoms that cross-substitute onto Tl site vacancies --- we use a semi-realistic dirty $d$-wave model based on empirically determined electronic structure and \emph{ab-initio} impurity potentials, following the methodology used in Refs.~\onlinecite{Ozdemir2022} and \onlinecite{Broun2024}. Band structure is taken from a tight-binding parameterization of ARPES measurements \cite{Plate2005,Peets2007b}, with doping dependence generated by rigid band shift. The ARPES-determined band structure is renormalized by a constant factor of 0.78 to make the cyclotron mass agree with the average value \mbox{$m^* = 5.2~m_e$} inferred in dHvA studies \cite{Vignolle2008,Bangura2010,Rourke2010}.  The impurity potential for the Cu cross-substituent, including shifts to site energies and hopping integrals, has been calculated using supercell density functional theory and tabulated in Ref.~\onlinecite{Ozdemir2022}.  As in that work, we also include a small density of strong-scattering unitarity-limit impurities via a $t$-matrix term in the self energy. The strength of this term is parametrized by its contribution to the normal-state scattering rate,  $\Gamma_N^U = 0.2$~K, in line with previous work \cite{Lee-Hone2020,Ozdemir2022,Broun2024} and with the linear-to-quadratic cross-over temperature observed in superfluid density measurements on highly overdoped Tl$_2$Ba$_2$CuO$_6$ \cite{Deepwell2013}.

We assume a separable form for the pairing interaction, $V_p d_\veck d_{\veckP}$, where $V_p$ parametrizes the pairing strength. The eigenfunction $d_\veck$ is an expansion in terms of the two lowest $d$-wave harmonics of the square lattice, \mbox{$d_\veck \propto \left[\cos(k_x a)-\cos(k_y a)\right]+ a_{20} \left[\cos(2 k_x a)-\cos(2 k_y a)\right]$,} where $a$ is the in-plane lattice parameter. The simple, near-circular form of the Tl$_2$Ba$_2$CuO$_6$ Fermi surface and the sparseness of the $d$-wave gap harmonics mean that the expansion maps tightly onto the more familiar cylindrical form
$d(\phi) \propto \cos 2 \phi + \alpha \cos 6 \phi$, where $\phi$ is measured around the $(\pi,\pi)$ point. At each doping and temperature, the BCS gap equation is solved self-consistently in the presence of disorder, with $V_p$ adjusted as a function of doping and disorder to reproduce the Tl$_2$Ba$_2$CuO$_6$ $T_c(p)$ relation inferred from quantum oscillations  \cite{Bangura2010,Rourke2010} and plotted in Fig.~\ref{figA4}(a).  The end result of this procedure is a determination of the $d$-wave gap, $\Delta_\veck$, as a function of temperature, doping and disorder.  The process has been iterated for different choices of the second-harmonic parameter $a_{20}$, to find the solution that correctly reproduces the universal-limit thermal conductivity \cite{Lee1993}, $\kappa_0/T \approx (k_B^2/3 \hbar d) v_F/v_\Delta$. Here $v_F$ is the nodal Fermi velocity and the gap velocity $v_\Delta$ is the near-nodal momentum derivative of the $d$-wave gap along the Fermi surface. (In the self-consistent treatment, $v_\Delta$ is indirectly renormalized by impurity suppression of the gap magnitude.)  In experiments on $T_c = 15$~K Tl$_2$Ba$_2$CuO$_6$ \cite{Proust2002}, the residual thermal conductivity is observed to be  $\kappa_0/T = 1.41$~mW\,K$^{-2}$\,cm$^{-1}$, implying a velocity ratio $v_F/v_\Delta = 210$.  This constrains the magnitude of the $\cos 6 \phi$ term in the gap to be $\alpha = 12$\%.

Once the $d$-wave gap $\Delta_\veck$ is determined, the Sommerfeld specific-heat coefficient, $\gamma = dS/dT$, is calculated from the Bogoliubov quasiparticle entropy, which is a thermal average of the density of states, $N(\omega)$, as discussed in Ref.~\onlinecite{Hirschfeld1988}. $N(\omega)$ is obtained by solving for the renormalized frequency $\tilde\omega_\veck(\omega)$ and gap $\tilde\Delta_\veck(\omega)$ on the real frequency axis and carrying out a Fermi surface average:  $N(\omega) = 2 N_0\mathrm{Re}\big\langle\tilde \omega_\veck(\omega)/\sqrt{\tilde\omega_\veck^2(\omega) - \tilde\Delta_\veck^2(\omega)}\big\rangle_\mathrm{FS}$, where $N(0)$ is the single-spin density of states at the Fermi level.

\subsection{Inferences from the theoretical model}

The weak $T_{\rm c}$ dependence of the measured heat-capacity anomaly provides important constraints on the origin of the suppression of superconductivity in the strongly overdoped regime. As discussed in the main text, the doping dependence of the anomaly is well reproduced by assuming a  fixed level of realistic cation disorder while allowing the pairing strength to vary with doping. An alternative possibility is that the pairing interaction remains approximately constant, with the observed decrease of $T_{\rm c}$ arising predominantly from increasing disorder level. The latter scenario, shown in Fig.~\ref{figA3}, predicts a $40\%$ reduction in $\Delta\gamma(T_{\rm c})/\gamma_{\rm n}$ over the range studied in this work. This is inconsistent with the experimentally observed weak dependence of the anomaly on $T_{\rm c}$, showing that the observed doping dependence of $T_{\rm c}$ cannot be driven predominantly by an increase in disorder.

In addition, we can use the model to estimate how the pairing strength evolves with doping. In weak-coupling BCS theory, pairing strength is parametrized by the dimensionless product $N(0)V_p$, which in turn depends logarithmically on the characteristic energy of the exchange bosons.  For concreteness, we assume an energy scale of 100~meV for the characteristic spin-fluctuation frequency in Tl$_2$Ba$_2$CuO$_6$ \cite{Prelovsek2006}.
 As shown in Fig.~\ref{figA4}, the inferred pairing strength decreases smoothly with hole doping in the fixed-disorder scenario.  It remains finite at the edge of the superconducting dome, reaching a value approximately half that at optimal doping.

%% file: supplement.tex
\onecolumngrid
\section*{Supplementary Materials}
\setcounter{figure}{0}
\renewcommand{\thefigure}{S\arabic{figure}}

\begin{figure*}[h!]
    \centering
    \includegraphics[scale=0.82]{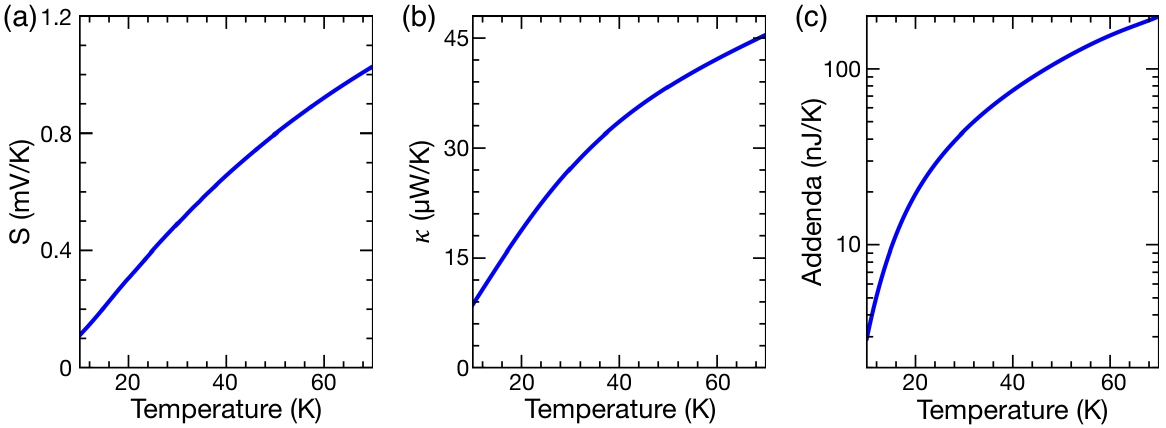}
    \caption{{\bf Characterization of the XEN-39398 calorimeter chip.} (a) Effective thermopower $S$, (b) thermal conductance $\kappa$, and (c) addenda heat capacity measured in this work.}
    \label{figS1}
\end{figure*}

\begin{figure*}[h!]
    \centering
    \includegraphics[scale=0.82]{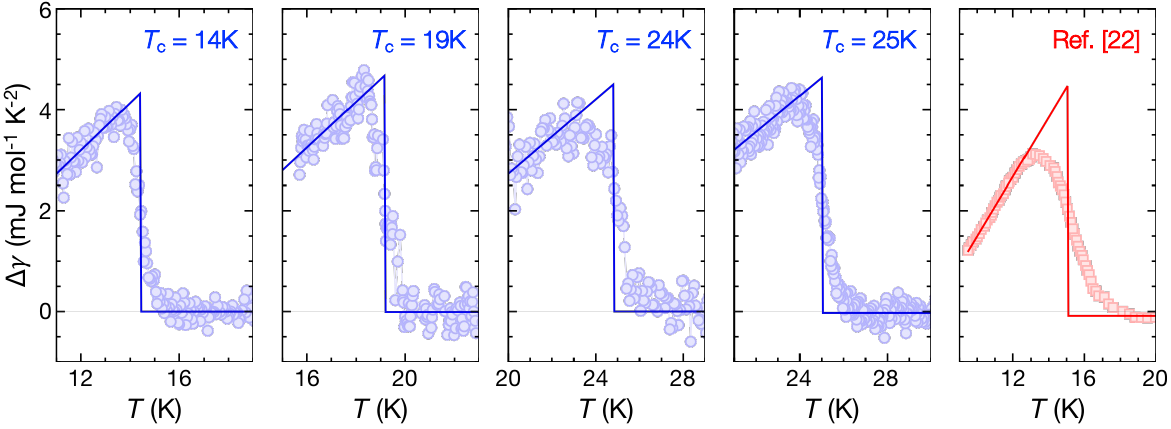}
    \caption{{\bf Entropy balance fits on all datasets.} Fits to the background-subtracted specific heat data prior to any smoothing, used to obtain the intrinsic jumps discussed in the main text. A similar analysis of the digitized data from Ref.~\cite{Carrington1996} is shown for comparison.}
    \label{figS2}
\end{figure*}

\begin{figure*}[h!]
    \centering
    \includegraphics[scale=0.82]{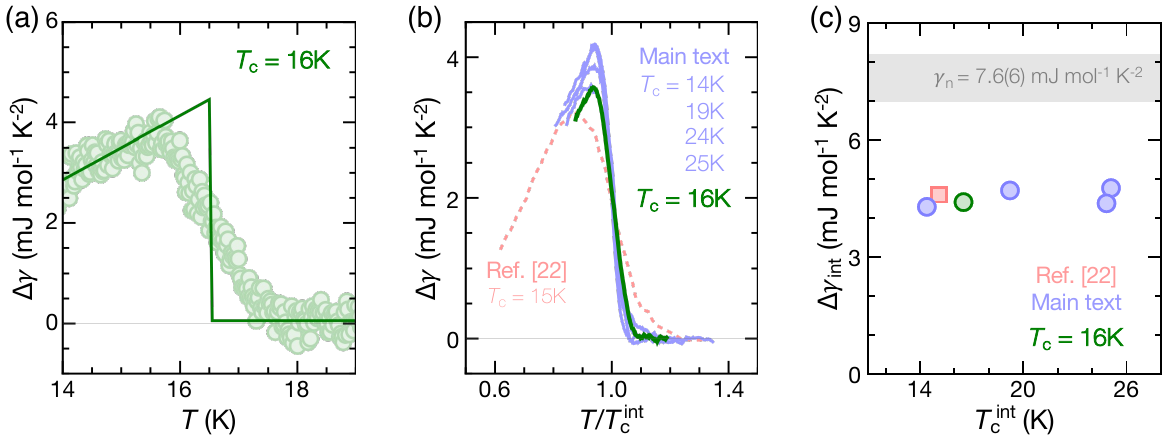}
    \caption{{\bf Additional supporting dataset.} (a) Electronic specific heat of a 4.8\,\textmu{}g crystal from Dresden, annealed to $T_{\rm c}=16$\,K using a different annealing protocol (1\,bar O$_2$, 380$^\circ$C for 5\,days). An entropy balance fit of the dataset is shown for reference.
    (b) Smoothed data, plotted against reduced temperature and overlaid with the datasets in Fig~\ref{fig2}c, showing good agreement in the thermodynamics of all far-overdoped datasets. 
    (c) Intrinsic $\Delta\gamma_{\rm int}$ obtained from an entropy balance fit in panel a, showing quantitative agreement with the analysis presented in the main text.}
    \label{figS3}
\end{figure*}

\begin{table*}[h!]
\caption{Details of Tl$_2$Ba$_2$CuO$_6$ single crystals measured and analyzed in this work.} \label{table1}
\begin{ruledtabular}
\begin{tabular}{cccccc}
{Dataset} & {Sample source} & {Thermal mass} & {Annealing conditions} & {$T_{\rm c}$} & {Polynomial fit range}\\
\colrule
A1-od14K & Cambridge & 2\,\textmu{}g & 1\,bar O$_2$ & 14\,K & 18--24\,K\\
A1-od25K & Cambridge & 2\,\textmu{}g & 0.01\,bar O$_2$ & 25\,K & 27--32\,K\\
D1-od19K & Dresden & 43\,\textmu{}g & 1\,bar O$_2$ & 19\,K & 22--28\,K\\
D4-od24K & Dresden & 25\,\textmu{}g & 0.1\,bar O$_2$ & 24\,K & 26--32\,K\\
D2-od31K & Dresden & 30\,\textmu{}g & 0.01\,bar O$_2$ & 31\,K & 35--45\,K\\
D4-od44K & Dresden & 25\,\textmu{}g & 0.001\,bar O$_2$ & 44\,K & 50--60\,K\\
C1-od47K & Bristol & 4.2\,\textmu{}g & 0.003\,bar O$_2$ & 47\,K & 52--72\,K\\
\end{tabular}
\end{ruledtabular}
\end{table*}